# Analyzing Exoplanet Phase Curve Information Content: Toward Optimized Observing Strategies

Ben Placek

*Department of Sciences, Wentworth Institute of Technology, Boston, MA 02115*

Daniel Angerhausen[1] [2]

*Center for Space and Habitability, University of Bern, Sidlerstrasse 5, 3012 Bern, Switzerland*

Kevin H. Knuth [3]

*Department of Physics, University at Albany (SUNY), Albany, NY 12222*

## ABSTRACT

Secondary eclipses and phase curves reveal information about the reflectivity and heat distribution in exoplanet atmospheres. The phase curve is composed of a combination of reflected, and thermally emitted light from the planet, and for circular orbits the phase curve peaks during the secondary eclipse or at an orbital phase of 0.5. Physical mechanisms have been discovered which shift the phase curve maximum of tidally locked close in planets to the right, or left, of the secondary eclipse. These mechanisms include cloud formations, and atmospheric superrotation, both of which serve to shift the thermally bright hot-spot, or highly reflective bright spot, of the atmosphere away from the sub-stellar point. Here we present a methodology for optimizing observing strategies for both secondary eclipses and phase curves with the goal of maximizing the information gained about the planetary atmosphere while minimizing the (assumed) continuous observation time. We show that we can increase the duty cycle of observations aimed at the measurements of phase curve characteristics (Amplitude, Phase offset) by up to 50% for future platforms such as CHEOPS and JWST. We apply this methodology to the test cases of the Spitzer phase curve of 55-Cancri-e, which displays an eastward shift in its phase curve maximum, as well as model-generated observations of an ultra-short period planet observed with CHEOPS.

*Subject headings:* astronomical databases: miscellaneous — methods: data analysis — methods: statistical — planetary systems — planets and satellites: detection — planets and satellites: fundamental parameters — planets and satellites: general — techniques: photometric — planets and satellites: individual (55-Cancri-e)

[1]NASA Goddard Space Flight Center, Exoplanets & Stellar Astrophysics Laboratory, Code 667, Greenbelt, MD 20771

[2]Blue Marble Space Institute of Science, 1001 4th Ave, Suite 3201, Seattle, Washington 98154

[3]Department of Informatics, University at Albany (SUNY), Albany, NY 12222

## 1. Introduction

Exoplanets in edge-on orbital configurations enable us to observe three key events: the transits and eclipses when the planet passes in front of and behind the star, respectively, as well as the out of eclipse time series - the phase curve. Comprised of time-varying thermal emission or reflected light, phase curves reveal information about the temperature and reflectivity of the exoplanet atmo-spheres depending on the observational bandpass. This type of observation opened up an entirely new field for exoplanet characterization starting with the first thermal phase curves obtained by the *Spitzer* Space Telescope (*Spitzer*) (Harrington et al. 2006; Knutson et al. 2007). It is now possible to conduct comparative studies of exoplanet phase curves on sizable samples obtained from the *Kepler* Space Telescope (*Kepler*) (Esteves et al. 2013; Heng and Demory 2013; Angerhausen et al. 2015). Additional information about exoplanets can be gained by using multi-color observations (Shporer et al. 2014; Placek et al. 2016) or low resolution (R $\sim$ 10-100) spectroscopic phase curves (Stevenson et al. 2014) for an in-depth analysis of single systems.

Detailed studies of phase curves can be observationally expensive as they require one to monitor the planet over the course of an entire orbit. Therefore it would be useful to determine if one could constrain atmospheric properties while minimizing the observation time in hopes of securing time on increasingly competitive telescopes, and performing comparative studies of the atmospheres of short period planets. This is especially true when observing planets with orbital periods more than a few days. This work was inspired by that of Loredo et al. (2012), where Bayesian experimental design was used to determine the most informative portions of radial velocities to observe. Krick et al. (2016) recently published a similar novel method for sparsely sampling exoplanet phase curves, which better align with the Spitzer observing schedule and reduce observation time. Here, using Bayesian model selection, we present a study of how one can derive certain phase curve characteristics while minimizing the observation time centered around the secondary eclipse, as well as a discussion on the most important portions of the phase curve for constraining the peak offset associated with cloud formations, and atmospheric superrotation.

In section 2 we describe in detail the methods and models used in our study and show in section 3 how we used them in our analysis. Section 4 illustrates an exemplary application to data from the *Spitzer* space telescope. Section 5 demonstrates this methodology in the context of the CHaracterising ExOPlanets Satellite (CHEOPS). In section 6 we summarize and discuss our results.

## 2. Methods

This section will explore the two phase curve models used in the analysis as well as the adopted Bayesian framework, prior probabilities, and likelihood function assignments.

### 2.1. Forward Models

For this analysis we focus on two forward models; one that models exoplanet phase curves and secondary eclipses while allowing for offsets in the maximum of the phase curve, and another model that neglects this offset. A shift in the maximum is indicative of the presence of atmospheric superrotation or cloud cover, and can appear on either side of the secondary eclipse. With tidally locked short-period planets, the phase curve has been observed to peak outside of secondary eclipse, which indicates that a bright-spot is shifted from the substellar point on the planet. Here we are not concerned with what is causing the shift, only our ability to detect it. For simplicity, we use *Kepler*-7b and 55-Cancri-e as illustrative examples as they both show bright-spot offsets, and the additional photometric effects known to be associated with exoplanet light curves such as ellipsoidal variations, and Doppler beaming (Placek et al. 2014, 2015; Loeb and Gaudi 2003; Angerhausen et al. 2015; Esteves et al. 2013; Faigler and Mazeh 2011; Shporer et al. 2011; Jackson et al. 2012; Mazeh et al. 2011; Mislis and Hodgkin 2012; Mislis et al. 2011; Pfahl et al. 2008) are small enough to be neglected. (Demory et al. 2013, 2016).

We approximate the phase curves as a sinusoid given by:

$$F(t) = F_{\text{ecl}} + (F_{\text{max}} - F_{\text{min}})\cos(\theta t + \delta\phi) + F_{\text{min}} \quad (1)$$

Several examples of the model light curves are shown in Figure 1. Notice as the shift in the bright spot increases from the substellar point, the phase curve maximum shifts away from an orbital phase of 0.5. Four parameters are used to describe the two models: phase curve amplitude ($F_{\text{max}}$), phase curve minimum ($F_{\text{min}}$), secondary eclipse depth ($F_{\text{ecl}}$), and the bright spot shift ($\delta\Phi$). The bright spot shift is set to zero for the model that neglects a shift in the phase curve and thus requires only the phase curve amplitude, and secondary eclipse depth. Since the goal is to minimize observation time, one may not be able to precisely constrain the night-side temperature as it depends on the ability to estimate $F_{\min}$. However, even for full phase curves the night-side fluxes aren't well constrained. This will be discussed further in Section 3.3. Secondary eclipses are generated using the method of Mandel and Agol (2002). One may also make use of more sophisticated models such as the longitudinal mapping method (Demory et al. 2016), which partitions the atmosphere into panels of varying brightness.

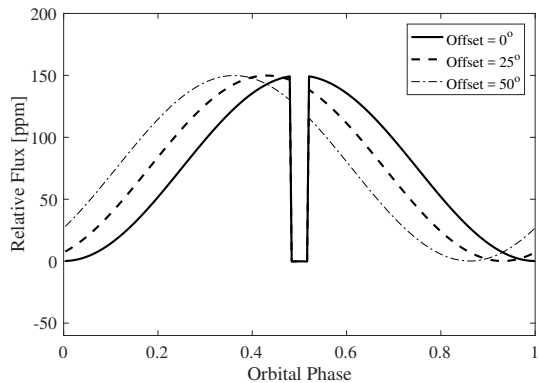

Fig. 1.— Morphologies of exoplanet phase curves with increasingly shifted phase curve peak for offsets of $0^o$, $25^o$, and $50^o$. The secondary eclipse occurs at 0.5. in orbital phase.

For the remainder of this paper we will refer to the two models as follows:

$$M_{\text{offset}} = \text{phase curve} + \text{phase offset}$$
$$M_{\text{no-offset}} = \text{phase curve}$$

## 2.2. Phase Curve Entropy Maps

With the goal of determining how many observations are required to constrain certain effects, it is also important to identify the parts of the phase curve that yield the most information concerning said effects. This can be accomplished by considering the principle of maximum entropy. From an information theoretic standpoint, entropy $H$, quantifies a lack of information. For a discrete set of probabilities $p_i$, the entropy, or Shannon Entropy is given by

$$H = -\sum_{i=1}^{N} p_i \log p_i. \qquad (2)$$

The location on the phase curve which will yield the most information about a set of observables corresponds directly to the locations where the entropy is a maximum (i.e. where the observer knows the least). In order to compute the entropy as a function of orbital phase $H(\phi)$, we must first define any prior knowledge about the system. The following sections will illustrate three simple cases of prior knowledge, and the resulting entropy.

### 2.2.1. Completely Uninformed

In the case where the observer is completely uninformed concerning the parameters of interest (see Table 1), the prior information about the phase curve can be quantified using uniform priors over reasonable ranges. Figure 2-A displays an "Entropy-map" for the completely uninformed situation. Bright regions are areas of high entropy, whereas dark regions are areas of low entropy. In this particular situation, one can see that the entropy is maximized during secondary eclipse. There is also a significant amount of entropy during ingress and egress of the secondary eclipse, but very little near the ingress/egress of the transit.

### 2.2.2. Known Secondary Eclipse Depth

In some situations, observations of the secondary eclipse may have been made previously and the eclipse depth known to some precision. If this were the case, one might place a Gaussian prior on the eclipse depth according to the previously measured value, and uniform priors on the remaining parameters. This results in an "Entropy-map" that peaks on the wings on the phase curve (Figure 2-B). It should be noted that there are two peaks since the phase offset is not known. Even in this situation, the secondary eclipse incorporates a high amount of entropy.

### 2.2.3. Known phase offset

The final illustrative example is that which would occur if the phase offset was measured previously. For instance, there seems to be a direct correlation with the equilibrium temperature of a planet and which direction (eastward/westward of the sub-stellar point) the phase curve is shifted (Esteves et al. 2015). Here, one could assign a Gaussian distribution for the phase offset, while assigning uniform probabilities for the remaining parameters. Figures 2-C and -D display the "Entropy-maps" for an westward shift, and a eastward shift, respectively. Again the wings of the phase curve seem to be important pieces of information. Note that the regions of the phase curve where entropy is high indicate locations where one would expect to gain the most information from further observations.

### 2.2.4. Importance of the secondary eclipse

The secondary eclipse is undoubtedly one of the most informative features of a transiting planet's phase curve. With broadband photometry, the secondary eclipse can yield information on the planetary albedo and/or day-side temperature. When observations are made over many spectral windows, even more information can be gleaned from the secondary eclipse alone. Together with the fact that in the most uninformative situation, the entropy is maximized during secondary eclipse. Therefore, for the remainder of this study we assume that all observations will be centered around the middle of the secondary eclipse.

## 2.3. Bayesian Model Selection

Bayesian model selection relies on one's ability to estimate the Bayesian odds ratio, $O$, which robustly compares two competing models using the Bayesian evidence. The evidence for a particular model represents the probability that that model correctly describes the observed data independent of the model parameters, and is computed by the following:

$$Z = \int P(\theta)P(D|\theta)d^N\theta. \quad (3)$$

Here, $D$ represents the observed data, $N$ the number of model parameters, $\theta$ the set of model parameters, $P(\theta|I)$ the prior probability for model parameters $\theta$, and $P(D|\theta, I)$ the likelihood function. Because the evidence is an integral over the entire parameter space, it acts to penalize models with large volumes in parameter space. In other words, the evidence has an inherent Occam's razor effect — all things being equal, simpler models are favored over complicated models.

In practice, the log-evidence is typically computed so that Bayesian model selection is performed using the log-odds ratio

$$\ln O = \ln Z_1 - \ln Z_2, \quad (4)$$

where $Z_1$ and $Z_2$ are the Bayesian evidences for models $M_1$ and $M_2$, respectively. Note that this assumes there is no reason to suspect that one model is favored over the other a-priori.

In general, there are varying degrees of detection, as any simulation yielding a log-odds ratio between $1.0 < \ln O < 2.0$ is deemed a positive detection for model $M_1$, between $2.5 < \ln O < 5.0$ represents a strong detection of $M_1$, and log-odds ratios greater than five are deemed overwhelmingly positive detections (von der Linden et al. 2014). A log-odds ratio less than zero indicates a favorability for model $M_2$ over $M_1$. For a more detailed treatment of Bayesian evidence refer to (Knuth et al. 2015; Placek et al. 2014, 2015, 2016; Sivia and Skilling 2006; von der Linden et al. 2014).

### 2.3.1. Priors and Log-likelihood Function

In order to estimate the Bayesian log-evidence for each model, one needs to define prior probabilities for each model parameter, and a likelihood function.

Prior probabilities represent one's knowledge of a system before having analyzed any data. For this analysis, it is assumed that nothing is known a-priori about any of the model parameters used to describe the phase curve/offset. Therefore, uninformative uniform distributions over reasonable ranges of parameter values are chosen to describe our prior knowledge of the light curve. These are displayed in Table 1.

The likelihood function is the probability of observing the data given a set of model parameters. It depends on the forward model, and the expected nature of the noise. For this analysis it is assumed that the noise is Gaussian in nature so that the log-likelihood function takes on the form

$$\log L = -\frac{1}{2\sigma^2}\sum_{i=1}^{N}(F(\phi_i) - D_i)^2 - \frac{N}{2}\log 2\pi\sigma^2. \quad (5)$$

Here $\sigma^2$ is the noise variance, $F(\phi_i)$ the model flux at the $i^{th}$ orbital phase $\phi$ as given by equation (1), and $D_i$ is the $i^{th}$ observation.

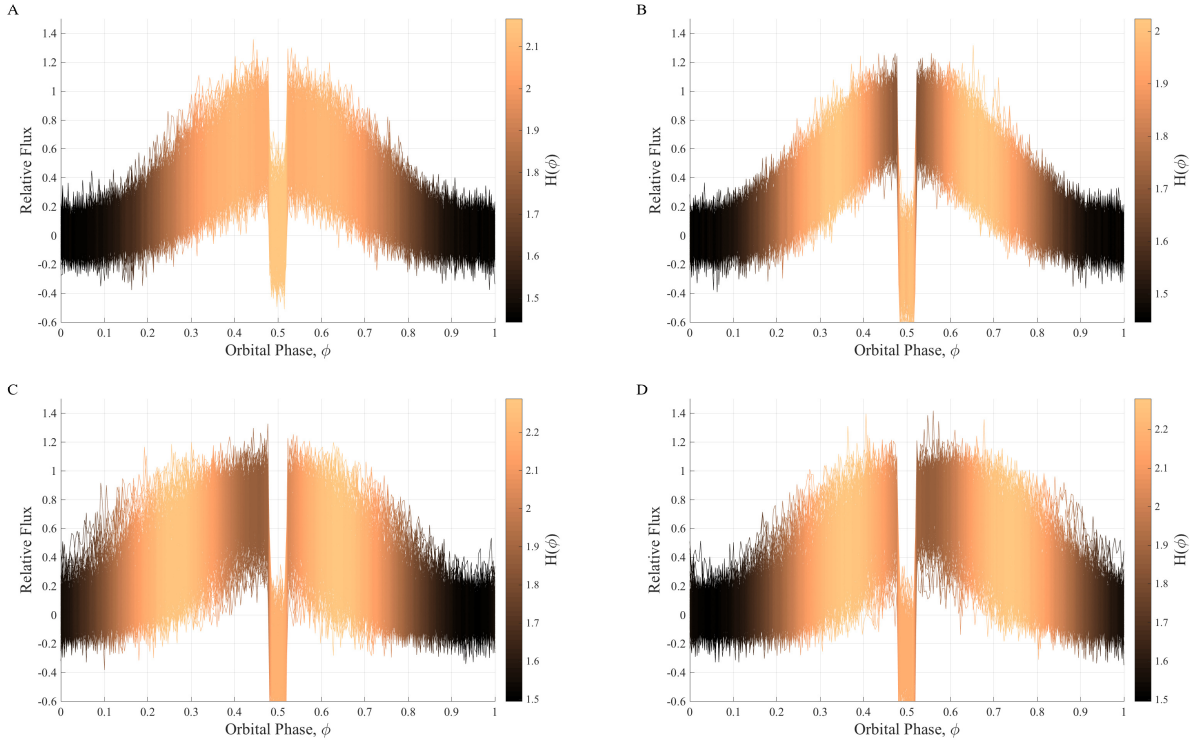

Fig. 2.— "Entropy Maps" of model generated phase curves. Each panel shows the locations of least information for the three cases described in Section 2.2.

## 3. Analysis

This section will summarize the synthetic data generation process, the results of applying Bayesian model selection to determine the amount of data required to detect a shift in the phase curve maximum, and the ability to constrain the magnitude of the shift by means of parameter estimation. All simulations were performed using the MultiNest algorithm (Feroz and Hobson 2008; Feroz et al. 2009, 2013), which is a variant on the Nested Sampling algorithm (Sivia and Skilling 2006).

Sections 3.1-3.3 outline a study of simulated observations of *Kepler*-7b, Section 4 investigates the amount of the phase curve necessary to constrain certain properties of the super-Earth 55-Cancri-e, and Section 5 investigates the outlook for future observations with the CHaracterising ExOPlanets Satellite (CHEOPS). The results of these analyses are summarized in Table 2

### 3.1. Model-Generated Data

As an illustrative test case, we assume a phase curve amplitude and secondary eclipse depth of 48ppm and 41.5ppm, respectively. These correspond to the estimated values for the low-density transiting hot-Jupiter *Kepler*-7b, which displays a phase offset of $\approx 40^o$ attributed to a cloud dominated western hemisphere, and does not display any substantial Doppler beaming, or ellipsoidal variations (Demory et al. 2013). The data were generated using the model described in Section 2.1 and Gaussian additive noise with a standard deviation of 20ppm, just slightly better than the photometric precision of K2. In order to generate model phase curves, additional parameters corresponding to *Kepler*-7b were assumed such as the orbital period $P = 4.8855$ days and inclination $i = 86.5^o$, planetary mass $M_p = 0.433 M_{\rm Jup}$ and radius $R_p = 1.478 R_{\rm Jup}$, stellar mass $M_\star = 1.347 M_\odot$ and radius $R_\star = 1.843 R_\odot$ (Latham et al. 2010). For illustrative purposes we take the ideal-ized case and assume that there is no systematic noise in the data. We also assume that only a single set of observations is made. That is, observations of certain orbital phases are not made over the course of many orbits of *Kepler*-7b and subsequently stacked together.

| Parameter | Variable | Distribution |
|---|---|---|
| Phase Curve Amplitude (ppm) | $F_{\max}$ | U(0,200) |
| Phase Curve Minimum (ppm) | $F_{\min}$ | U(0,100) |
| Sec. Eclipse Depth (ppm) | $F_{\rm ecl}$ | U(0,200) |
| Bright Spot Offset (deg) | $\delta\phi$ | U(-90,90) |

Table 1: Prior Distributions for phase curve and secondary eclipse model parameters. U signifies a uniform distribution over the range inside the parentheses.

Since one's ability to detect the bright spot offset will likely depend on the magnitude of the shift, which is unknown a-priori, we will look at five different shifts from $20^o$ to $50^o$. Also, phase curve maxima have been observed on both sides of the secondary eclipse. Therefore, since we do not a-priori know which side the maximum will appear, the out-of-eclipse data is incorporated symmetrically around ingress and egress. In order to determine how much data is necessary before and after the secondary eclipse, we generate 20 datasets each with an increasing amount of data outside of the eclipse starting with only observations during eclipse and increasing the amount out of eclipse by 0.01 in orbital phase. This process reaches out to 0.2 in orbital phase, which covers approximately 40% of the total phase curve. The cadence of the observations is taken to be 29.4 minutes, which corresponds to the long-cadence observing mode of *Kepler*. This translates to approximately 24 data points per 0.1 in orbital phase. Examples of this process are displayed in Figure 3.

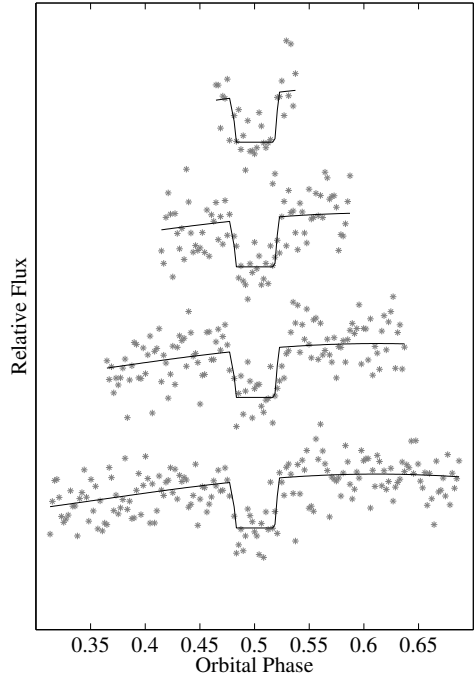

Fig. 3.— Examples of the model generated data used in this analysis. The top dataset incorporates only data during secondary eclipse, the second from the top includes observations out to 0.05 in orbital phase to either side of the secondary eclipse, the third from the top and bottom datasets include observations out to 0.10 and 0.15 in orbital phase, respectively. The amplitude of the phase curve and secondary eclipse depth were both taken to be 48ppm and 41.5ppm, respectively. The phase offset was assumed to be $60^o$.

### 3.2. Detecting the Shift

We define the detection of the phase offset to be when the model assuming a phase offset is favored over the non-shifted model in terms of the log-odds ratio. Specifically, when the difference in log-evidence exceeds five (overwhelming evidence). The estimated Bayesian log-evidences are displayed in Figure 4. The bottom x-axis in the figure represents the number of observations taken outside of secondary eclipse measured in orbital phase, whereas the top x-axis measures the number of observations in units of hours symmetrically to either side of the secondary eclipse.

Phase offsets of $40^o$ and $50^o$ are essentially detectable with only a few hours before and after secondary eclipse. However, the evidence for model $M_{\text{offset}}$ does not exceed the overwhelming status until after $\approx 11.8$ hours, and $\approx 29.4$ hours for shifts of $50^o$ and $40^o$, respectively. For phase offsets of $20^o$ and $30^o$, the log-odds ratios show that model $M_{\text{no-offset}}$ would be favored out to roughly 0.14 in orbital phase for a $30^o$ shift and 0.24 for a $20^o$ shift. This is indicated by the curves residing below zero out to those points. Observations of a $10^o$ phase offset did not yield a significant detection for model $M_{\text{offset}}$ out to approximately 0.7 in orbital phase ($\approx 82.1$ hours) outside of secondary eclipse. This will be discussed further in Section 3.4. It should be noted that these results are likely dependent on the signal to noise ratio and cadence of the observations. Increasing the amount of data, and/or obtaining more precise measurements will likely allow one to detect the phase offset with less out of eclipse data.

### 3.3. Constraining the Shift

In addition to detecting the phase offset, the ability to constrain the value of the offset is also important. For each simulation, the posterior distribution for the phase shift was obtained. The relative error, which we take to be the ratio of the standard deviation and the mean, for the phase offset for each simulation is displayed in Figure 5. For comparison, the relative error for the phase offset taking into account the entire phase curve is plotted as a horizontal dashed line. This horizontal dashed line is the benchmark as it takes into account the entire orbit. Notice that the relative error is within $\approx 5\%$ of that when the entire phase curve is taken into account at approximately 0.36 in phase for the $30^o$ offset, 0.3 in phase for the $40^o$ offset, and 0.08 in phase for the $50^o$ offset.

There also appears to be a small bump in each plot of the relative error indicating a region in the phase curve where the parameter uncertainties increase slightly. Figure 5 (right column) displays the region of the phase curve (vertical lines) corresponding to the locations of the bumps in relative error of the phase offset parameter. Each of the bumps correspond approximately to the location of the peak in the phase curve, which implies that the peak is important for constraining the offset. This also implies that it will be difficult to constrain the offset when the peak of the phase curve occurs during the secondary eclipse. Provided enough data, this would not be a problem with observations of the entire phase curve as the wings may reveal the presence of such a small shift. This will be discussed in more detail in Section 3.4.

Figure 6 displays the day and nightside flux estimates for the same 20 simulations. The left column shows that the precision of the dayside flux essentially plateaus at 0.1 in orbital phase outside of secondary eclipse for offsets of 30, 40, and 50 degrees. The right column shows the equivalent values for nightside flux. In particular, it shows that if given only secondary eclipse data (corresponding to zero on the axis labeled "Additional Orbital Phase") one cannot precisely constrain the nightside flux. This makes sense since there is no data to which the nightside flux is contributing. As more data is obtained the nightside flux is constrained more. Parameter estimates for the day and nightside fluxes may be dependent on the chosen model. Uncertainties estimated by the simple model assumed in this study may differ from those measured using a more sophisticated model that incorporates longitudinal mapping.

### 3.4. Limitations

In addition to the fact that results will vary depending on the characteristics of the desired observations (cadence, SNR, etc.), the relative errors in the phase offset parameter from Figure 5 imply another limitation to this methodology. As stated in the previous section, one's ability to estimate the magnitude of the phase offset depends on the location of the peak in the phase curve. This implies that for small peaks that occur during the secondary eclipse, it may be difficult to detect. For a planet with orbital period of 4.88d, the duration of the secondary eclipse will be long enough to hide the peak in the phase curve out to $\approx 8^o$, which would explain why it was difficult to detect the $10^o$ shift in Section 3.2. The top window in Figure 7 displays the peak to peak amplitude of the residuals obtained by subtracting models $M_{\text{offset}}$ and $M_{\text{no offset}}$ while changing both the phase offset and amount of data outside of sec-ondary eclipse. Notice that the residuals are very small (less than $\approx 5$ppm) for any phase offset less than $\approx 8^o$, which corresponds to the half-width of the secondary eclipse. The plot on the bottom of Figure 7 shows half of the secondary eclipse duration as a function of orbital period in units of degrees along the planet's circular orbit. The idea being that in order for the peak to occur outside of the secondary eclipse, the peak of the phase curve must be shifted by an angle greater than the angle subtended by the planet as it orbits over the course of half of the secondary eclipse. The secondary eclipse is indicated by the shaded region in the plot. This shows that the reason that phase offsets less than $10^o$ will not be detected is because the peak very nearly occurs during the secondary eclipse.

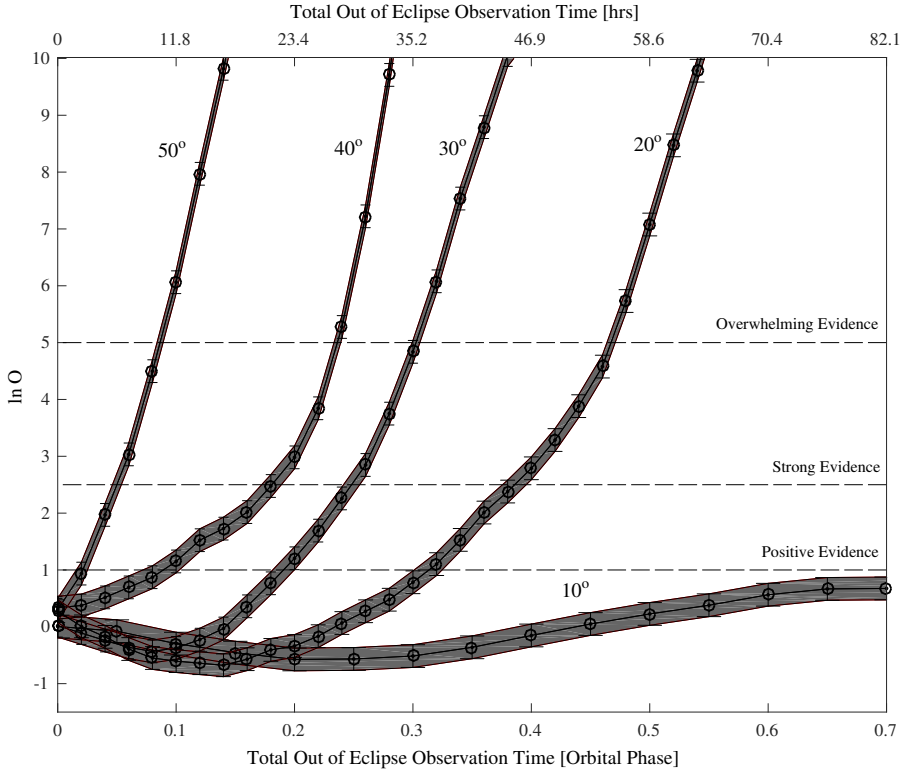

Fig. 4.— Log-odds ratios for simulated phase offsets of $20^o$, $30^o$, $40^o$, and $50^o$ as a function of out-of-eclipse observation time. The net out-of-eclipse observation time is measured in orbital phase (lower x-axis) and hours (upper x-axis) and is the sum of the out-of-eclipse observation time on both sides of the secondary eclipse. As expected, in order to detect the offset of the phase maximum one needs to observe longer for smaller offsets.

As mentioned previously, the model-generated example described above is the idealized case in that we assume no systematic noise. In practice, correlated noise will make this methodology more complicated especially if there are linear trends close to secondary eclipse as in the idealized case the slope of the phase curve prior to ingress and after egress are really what allow one to differentiate the two models ($M_{\text{offset}}$ and $M_{\text{no-offset}}$). A detailed knowledge of instrumental systematics such as the intra-pixel sensitivity variations known to be associated with *Spitzer* observations will be necessary to use this method in practice.

Lastly, the computation of the Bayesian evidence can be cumbersome as it is a multi-dimensional integral over the model parameter space. However, when aiming to make robust predictions about the capabilities of future observing platforms, particularly when it comes to phase curve observations, it may not be absolutely imperative to use the Bayesian evidence. There are approximations the to Bayesian evidence including the Bayesian information criterion (BIC), which is given by

$$BIC = -2\log \hat{L} + Nlog(n), \qquad (6)$$

and the Aikake Information Criterion (AIC) given by

$$AIC = -2\log \hat{L} + 2N, \qquad (7)$$

where $\hat{L}$ is the maximum likelihood value for the a given model, $N$ the number of model parameters, and $n$ the number of observations. While these approximations are much easier to compute than the full Bayesian evidence, they do come with their own assumptions including that correlations among model parameters are small.

## 4. Application to *Spitzer* Photometry

This section will outline the application of this optimization strategy to the *Spitzer* phase curve of 55-Cancri-e; a transiting Super-Earth with mass $M_p = 8.09 \pm 0.26 M_\oplus$, radius $R_p = 1.990^{+0.084}_{-0.080} R_\oplus$ (Tsiaras et al. 2016; Nelson et al. 2014), and day and night side temperatures of $T_d = 2700 \pm 270$K and $T_n = 1380 \pm 400$K, respectively (Demory et al. 2016). The phase curve of 55-Cancri-e displays an offset of $41° \pm 12°$ east of the substellar point, making it an ideal test case for this methodology. For simplicity, we use the reduced observations from Demory et al. (2016) obtained in the supplemental extended data tables. These data consist of 7087 data points (neglecting the transit) taken over the course of 8 continuous visits spanning 9 hours each.

Similar to the simulations in Section 3.1, twenty subsets of the data were made out of the entire phase curve. Starting with only data from the secondary eclipse, each dataset had additional out of eclipse data. This out of eclipse data corresponded to 0.025 in orbital phase to both sides of ingress and egress (corresponding to an average of 160 data points). Figure 8 displays the log-odds ratio as a function of additional out of eclipse data. In this case, the log-odds ratio yielded an overwhelming evidence ($\ln O > 5.0$) with only 0.05 (in orbital phase) of additional data to either side of the secondary eclipse. The reason that one can disentangle $M_\text{Offset}$ and $M_\text{No-Offset}$ with less data than the example in Section 3.1 is due to the higher number of observations.

Finally, Figure 9 shows the parameter estimates for each simulation. As expected it is difficult to constrain the parameters with only data inside of the secondary eclipse. The parameters seem to be better constrained to their literature values obtained from Demory et al. (2016) after data between 0.1 and 0.2 in orbital phase outside of secondary eclipse are added. This corresponds nicely to the point at which the log-odds ratio indicates that $M_\text{offset}$ becomes strongly favored in Figure 8. Therefore one would need 10-20% of the entire phase curve centered on the secondary eclipse. Taking into account the duration of the secondary eclipse of 55-Cancri-e ( 5% of the entire phase curve), this indicates that one can detect the presence of the phase curve offset, and recover uncertainties comparable with previous studies with only 15-25% of the entire phase curve.

## 5. Applications to Future Observing Platforms

The next generation of observing platforms will be highly competitive when it comes to applying for observation time. In order to determine the best observation strategy, one must consider observational parameters such as the measurement cadence, and photometric precision.

### 5.1. CHEOPS

The Characterising ExOPlanets Satellite (CHEOPS) is a 30cm space telescope with a 2018 projected launch date. It is planned jointly by the European Space Agency and Switzerland with the primary goal of characterizing previously discovered planets around bright host stars with ultra-high precision photometry (Fortier et al. 2014; Broeg et al. 2013). CHEOPS will have a measurement cadence of 1 minute, and is predicted to attain a photometric precision of 20ppm for a bright ($6 < V < 9$) star per six hours integration time, and 85ppm for a dimmer V = 12 star over 3 hours integration time. In it's Sun-synchronous orbit, CHEOPS will orbit the Earth once every 100 minutes. This orbit will result in periodic interruptions in ob-serving that can span between %20 − %50 of the 100 minute orbital period.

Model-generated data was created using the appropriate cadence and photometric precision, assuming Gaussian distributed noise. Using the predicted numbers listed above, we calculate the uncertainty of an individual observation to be (20ppm) × $\sqrt{360} \approx$ 380ppm. Gaps in the data were also created every 20 minutes to simulate the Earth eclipse, Sun, etc..

The interruptions that will afflict CHEOPS will likely affect planets on varying scales, which depend on the orbital period of the planet. Planets with longer orbital periods should in theory be less affected by these interruptions as smaller portions of their phase curve will be missed throughout the interruptions. A single interruption will mask approximately 0.3% of the phase curve for a planet with an orbital period of 5d similar to *Kepler-7b*. We focus on observing an ultra-short period planet with an orbital period 1d similar to WASP-43b (Hellier et al. 2011) or KELT-16 (Oberst et al. 2017), where a single interruption will mask up to 2% of the phase curve. Both of these planets will be prime targets for observation with JWST and CHEOPS.

Figure 10 displays the process through which the simulated observations were generated. Each row adds two CHEOPs orbits worth of observations; one to either side of the secondary eclipse starting with four orbits and ending with twelve. Black points represent the raw observations, yellow circles are binned to 7.5 minutes, and the black curve represents the model used to generate the observations. Additionally, we vary the number of planetary orbits on which to observe.

After the synthetic data were generated, a series of simulations were conducted to create the 7 × 5 grid in Figure 11. Each pixel in Figure 11 represents a series of 20 simulations exactly analogous to those carried out in Figure 8. The color of each pixel represents the amount of observation time required for the Bayes' factor to exceed the "Strong evidence" designation (i.e. where $M_{\text{offset}}$ is strongly favored over $M_{\text{no offset}}$). Thus the darker regions of the grid indicate the need for longer observing runs in order to disentangle the two models.

### 5.1.1. Observing Ultra-short Period Planets

The advantage to observing ultra-short period planets is that a fewer number CHEOPS orbits will be required to observe the entire phase curve. However, the interruptions will also mask a larger portion of the phase curve. For a planet with a 0.8 day orbital period, approximately 12 CHEOPS orbits will cover the entire phase curve, and each interruption will mask approximately 1.8% of the phase curve. Figure 11 displays the results from the simulations varying the amplitude of the phase curve and offset of the peak as described in Section 5.1. The first slice in Figure 11 assumes that one only observes a single orbit of the planet continuously with the observations centered on the secondary eclipse. Each color represents the number of CHEOPS orbits required to detect the phase offset per observing run. The second slice assumes that one conducts two separate observing runs, and the third slice assumes one conducts four separate observing runs. The total number of CHEOPS orbits needed to detect the phase curve offset is equal to the number of observing runs multiplied by the number of CHEOPS orbits per observing run.

If the phase curve amplitude and offset are known a-priori, we can use Figure 11 to predict how much of the phase curve it is necessary to observe. Assuming that only a single continuous observation is made of a planetary orbit, and the amplitude and phase offset are known to be larger than 50ppm and $20^o$, respectively, we expect that the offset is detectable with 83 − 90% of the entire phase curve. For phase curves with amplitudes and offsets greater than 90ppm, and $30^o$, we expect to detect the offset with as much as 67 − 75% of the entire phase curve. Finally, for combinations of amplitude and offset greater than 130ppm and $40^o$ it can be expected to need only 33 − 42% of the phase curve. If it is possible to make two separate continuous observations of the planet, both of which centered around consecutive secondary eclipses, we can expect to detect phase amplitude/offset combinations greater than 70ppm and $30^o$ with 67 − 75% of the phase curve. For combinations of 110ppm and $40^o$ only 42−58% of the phase curve is needed. Lastly, if it is possible to make three separate continuous observations of the planet, all centered on consecutive secondary eclipses, amplitude and offset combina-tions greater than 110ppm and $30^o$ require only $50-58\%$ of the full phase curve, and combinations greater than 130ppm and $40^o$ require only $33-42\%$.

## 6. Conclusions

Observation time on future space telescopes will be extremely competitive, and it may be difficult to obtain observations spanning the entire phase curve of exoplanets with orbital periods longer than a day. The methodology presented here illustrates how one can make informative inferences on parameters of interest (day- and night-side flux, and phase offset) with less than the full phase curve. For a model-generated planet similar to *Kepler*-7b, assuming an ideal scenario with zero red noise we show that the fluxes and phase offset should be detectable if the offset is greater than $\approx 10^o$, with as much as 50% of the phase curve observed. If the planet displays a phase offset of $50^o$, it can be detected with as little as 10% of the entire phase curve. The posterior parameter estimates for the three parameters of interest were investigated to determine if one could attain a similar level of precision as when the entire phase curve is analyzed.

This strategy was applied to the *Spitzer* phase curve of 55-Cancri-e. We show that the phase offset is detectable with as little as 10-20% of the total phase curve, and parameter uncertainties become comparable to those made with the entire phase curve with between 15-20% of the total centered around the secondary eclipse.

Finally, we provide insight into the types of phase curves that should be detectable in the context of the Characterising ExOPlanets Satellite using 2-dimensional grids of Bayes' factors. In the case of an ultra-short period planet for which the entire phase curve can be observed with 12 CHEOPS orbits, approximately 40% of the parameter space yield detectable combinations of phase offset and phase curve amplitude. This detectable region of the parameter space expands up to 55% if four separate planetary orbits are observed with 10 CHEOPS orbits each. A smaller region of the parameter space concentrated between phase curve amplitudes of 130-150ppm and offsets of 40-50 degrees are detectable with between 4-7 CHEOPS orbits for a single planetary orbit. If four planetary orbits are observed, this region expands to between amplitudes of 90-150ppm and offsets of 30-50 degrees. If the amplitude and offset are known a-priori, we show that a wide range of combinations of phase amplitude and offset are detectable with less than the full phase curve. Extreme combinations greater than 130ppm and $40^o$ requiring only $\approx 35\%$ of the phase curve.

In addition to making phase curve observations more competitive when observation time is being allocated, this methodology would be particularly useful for population studies looking for correlations between day- and night-side temperatures, the phase offset, and other characteristics such as planetary radius, incident flux, stellar metallicity, etc. since a large number of planets will need to be observed.

## Acknowledgements

D.A. acknowledges the USRA NASA postdoctoral program and the support of the Center for Space and Habitability of the University of Bern. His work has been carried out within the frame of the National Centre for Competence in Research PlanetS supported by the Swiss National Science Foundation.

|  | *Kepler*-7b | 55-Cancri-e | CHEOPS Obs. |
|---|---|---|---|
| Platform | Model Generated | Spitzer | Model Generated |
| Photometric Precision [ppm] | 20 (per 30 min) | 364 (per 30s) | 20 (per 6 hr) |
| Cadence | 29.4 min | 30s | 1 min |
| # Data Points | 401 | 7087 | 960 (per cont. observing run) |
| Phase Offset [deg] | 10, 20, 30, 40, 50 | — | 10, 20, 30, 40, 50 |
| Phase Curve Amplitude [ppm] | 50 | — | 30, 50, 70, 90, 110, 130, 150 |
| Secondary Eclipse Depth [ppm] | 48 | — | 41.5 |
| Orbital Period [days] | 4.8855 | 0.7365499 | 0.8 |
| Minimum Detectability [% of total] | $> 70, 48, 30, 25, 10$ | 15 | See Sec. 5 & Fig. 11 |

Table 2: Summary of data used and results presented in this paper. In the case of the model generated data, the numbers listed in this table represent those that were assumed when creating said data. Listed are multiple phase offsets, which were analyzed in the case of *Kepler*-7b in Section 3.1. Additionally, multiple phase curve amplitudes are listed as both amplitude and offset were varied in the study of model generated CHEOPS data in Section 5. The minimum detectability represents the percentage of the entire phase curve needed to detect $M_{\text{offset}}$ in each case.

---

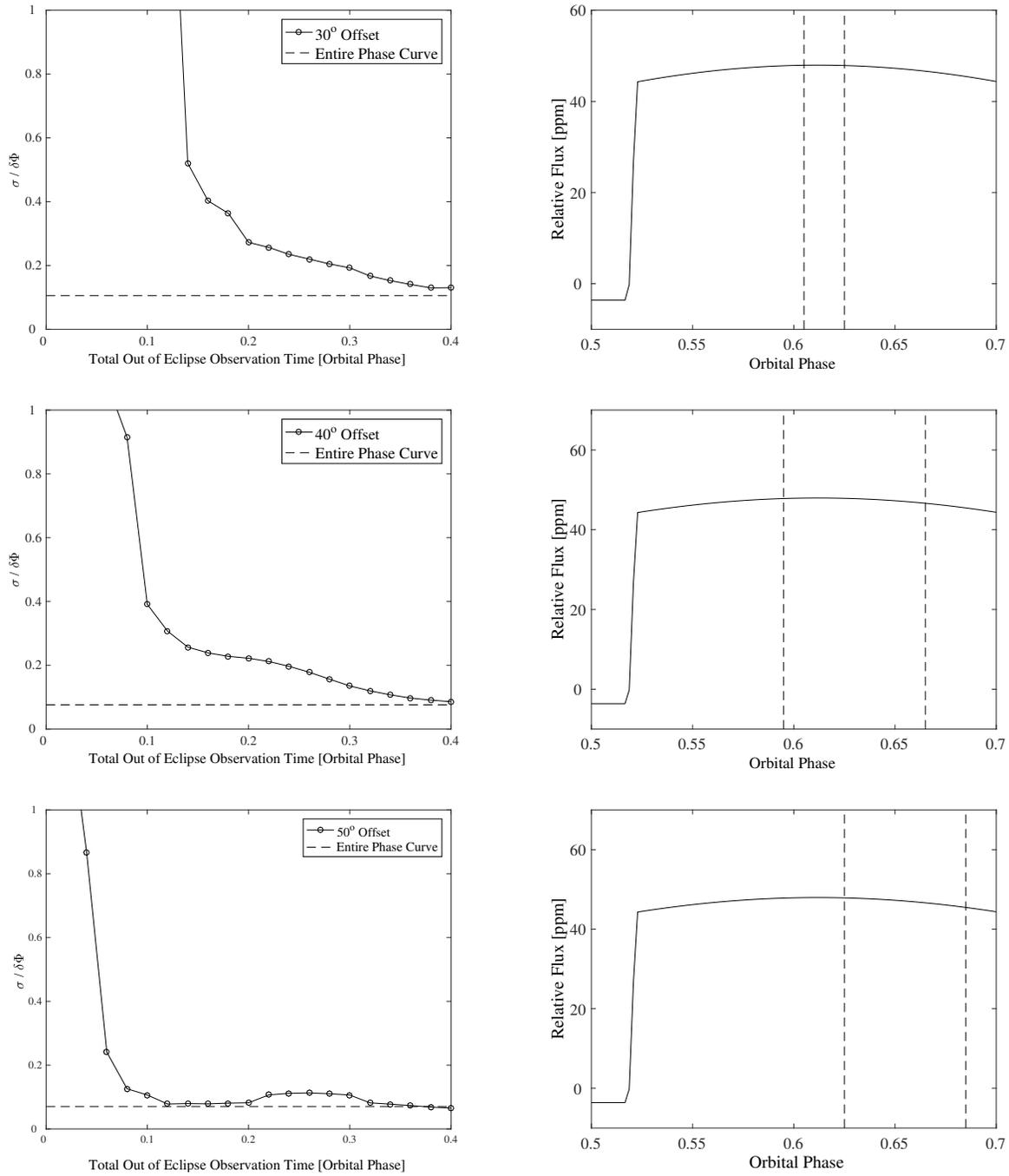

Fig. 5.— Left: Relative Errors for the offset in phase curve maximum for offsets of $30^o$ (top row), $40^o$ (middle row), and $50^o$ (bottom row). The horizontal dashed lines represent the relative errors for the offset if the entire phase curve was observed. Notice the bumps in each curve indicating a region of the phase curve which increases the uncertainty in the estimated offset angle. Right: Post secondary eclipse phase curves for each offset. The vertical lines represent to the time after secondary eclipse which corresponds to the bumps seen in the relative error figures. All of which occur near the peak in the phase curve maximum.

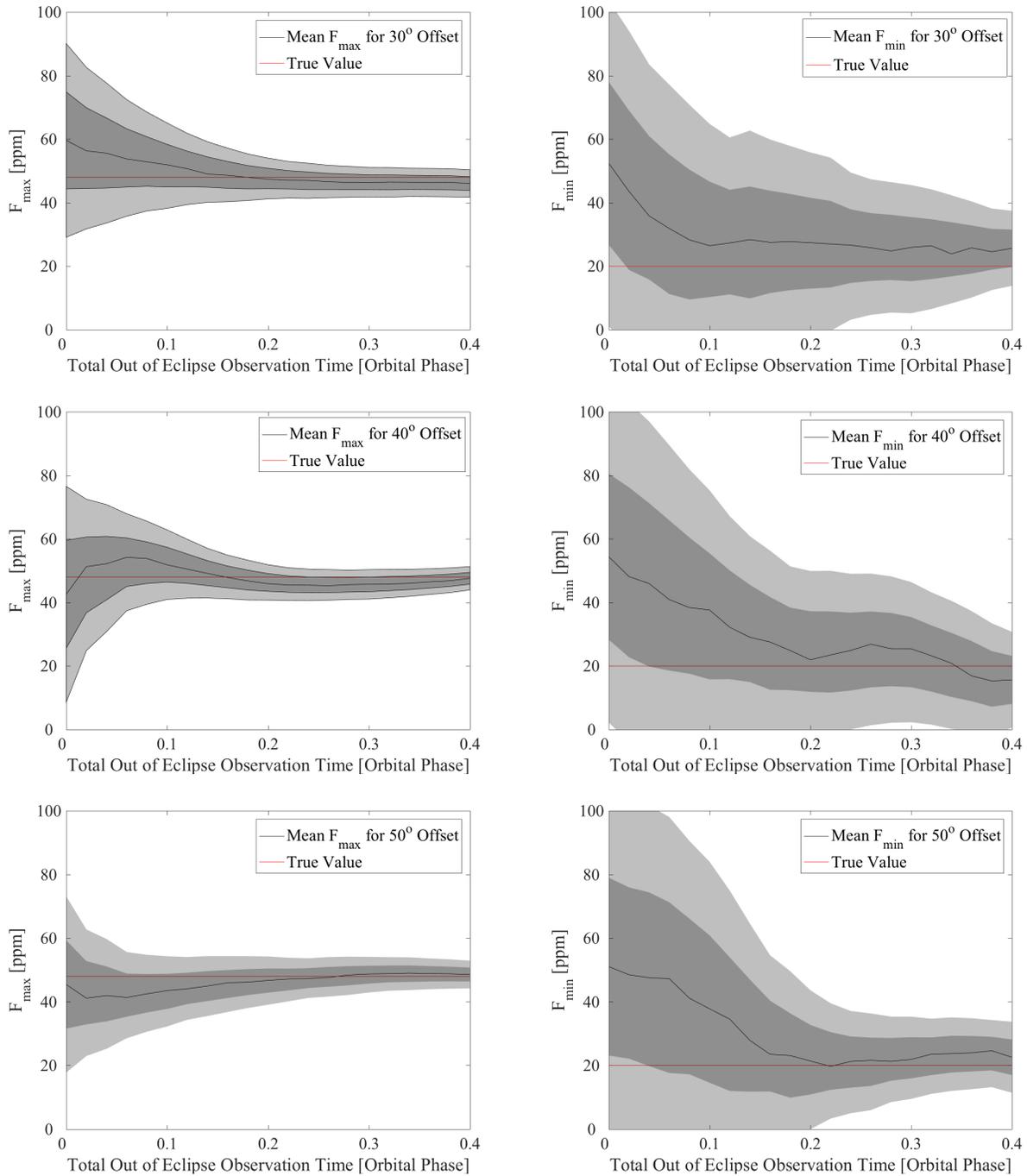

Fig. 6.— Dayside (left) and nightside flux ($F_d$ & $F_n$) estimates from the 20 simulations performed on simulated *Kepler*-7b data. Notice that the uncertainties in the nightside flux are very large when only the secondary eclipse or small amounts of data around the secondary eclipse are observed. More data yields more precise, and accurate results. The precision of the dayside flux measurements however plateau after approximately 0.2 in orbital phase outside of the secondary eclipse. The dark grey shaded region indicates the $1\sigma$ and $2\sigma$ limits, respectively.

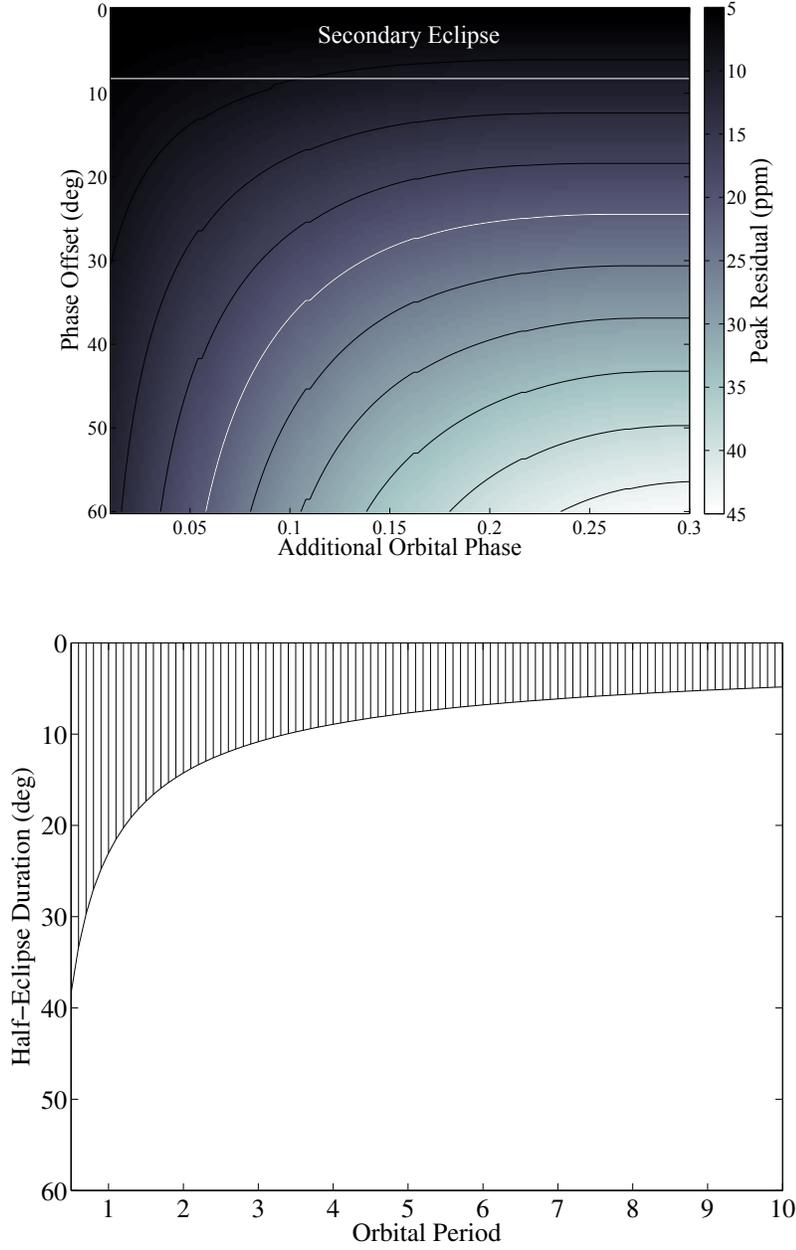

Fig. 7.— Top: Peak to peak amplitude of the difference between Model $M_{\rm offset}$ and Model $M_{\rm no\ offset}$ for varying degrees of phase offsets and amount of data outside of secondary eclipse. Lighter regions indicate larger amplitudes and the white contour represents the standard deviation of the additive Gaussian noise (20ppm) from simulations described in Section 3.2. Also labeled is the region inside which the phase curve maximum will occur during secondary eclipse. This implies that it will be difficult to detect any phase offsets less than $\approx 8^o$ ($\approx 0.1$ in terms of fraction of orbital phase). Bottom: Half of the secondary eclipse duration as a function of orbital period. Note that the eclipse duration will also depend on the radius of the host star, and impact parameter.

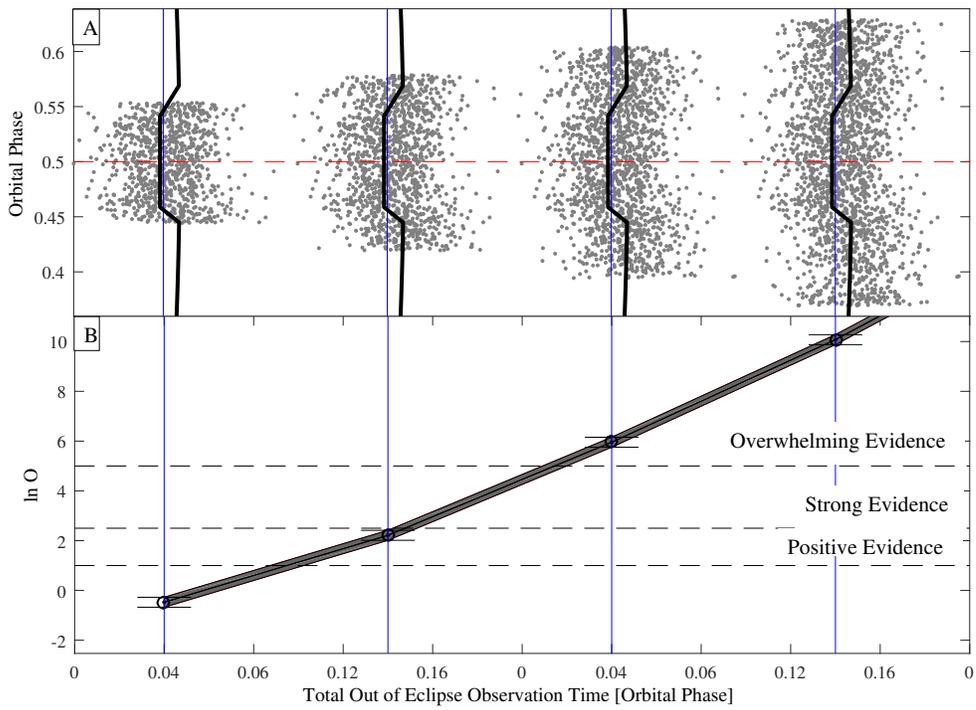

Fig. 8.— Log-odds ratios for Spitzer observations of 55-Cancri-e. The total out-of-eclipse observation time is measured in orbital phase (lower x-axis) and hours (upper x-axis) and is the sum of the out of eclipse observation time on both sides of the secondary eclipse.

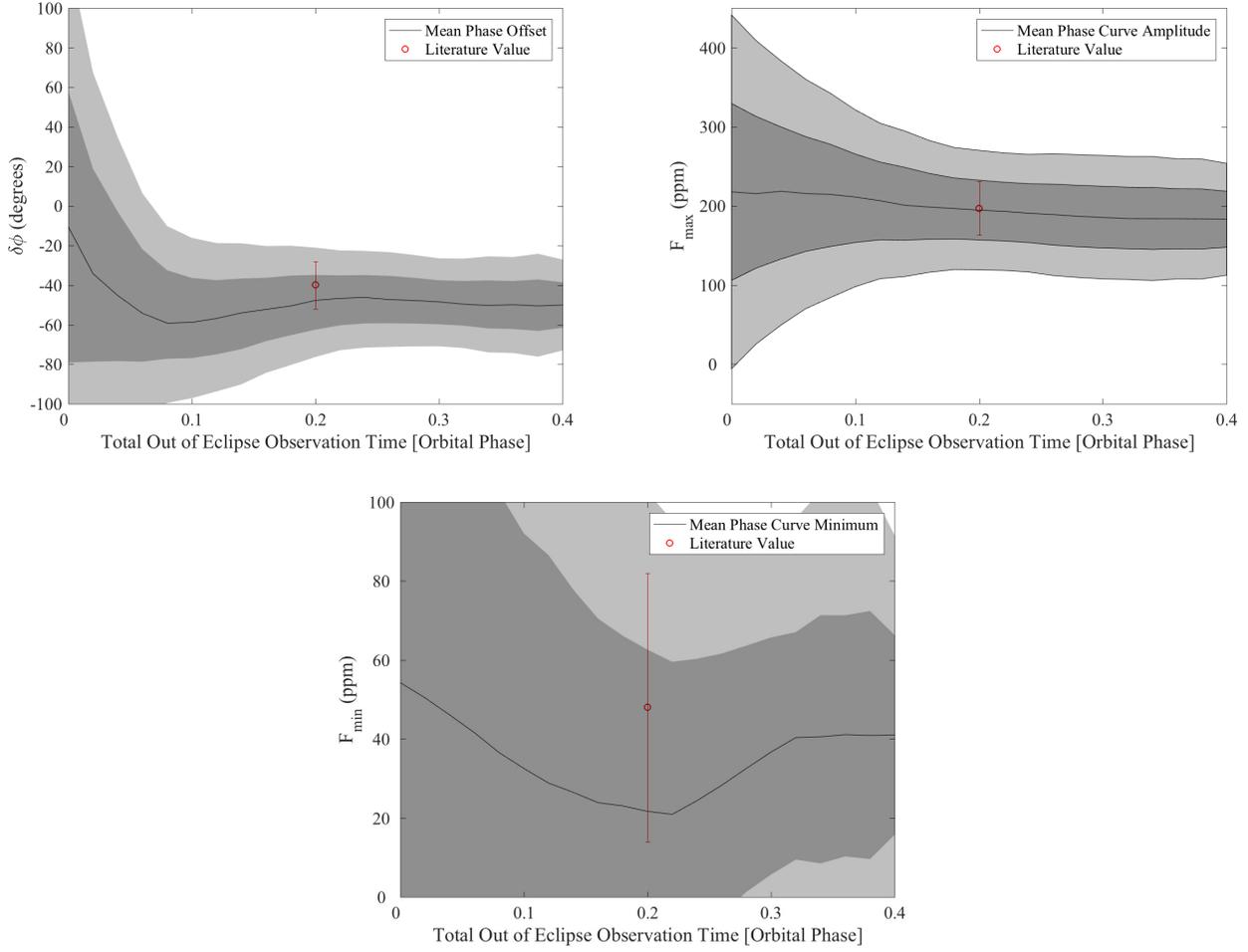

Fig. 9.— Parameter estimates for the simulations on the 55-Cancri-e phase curve (see Section 4). The precision and accuracy of the model parameters significantly improves when data is included out to 0.05-0.1 in orbital phase around the secondary eclipse. Literature values were taken from Demory et al. (2016) and are as follows: $\delta\phi = 40 \pm 12$ degrees, $F_{\max} = 197 \pm 34$ ppm, and $F_{\min} = 48 \pm 34$ ppm.

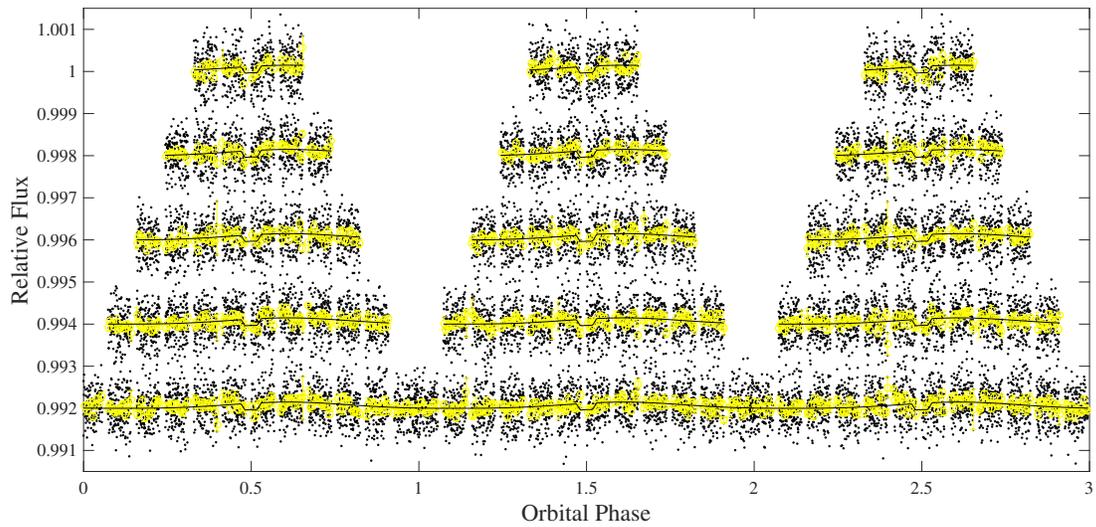

Fig. 10.— Simulated CHEOPS observations of an ultra-short period planet over three planetary orbital periods. Each row displays adds two CHEOPs orbits worth of observations; one to either side of the secondary eclipse starting with four orbits and ending with twelve. Black points represent the raw observations, yellow circles are binned to 7.5 minutes, and the black curve represents the model used to generate the observations.

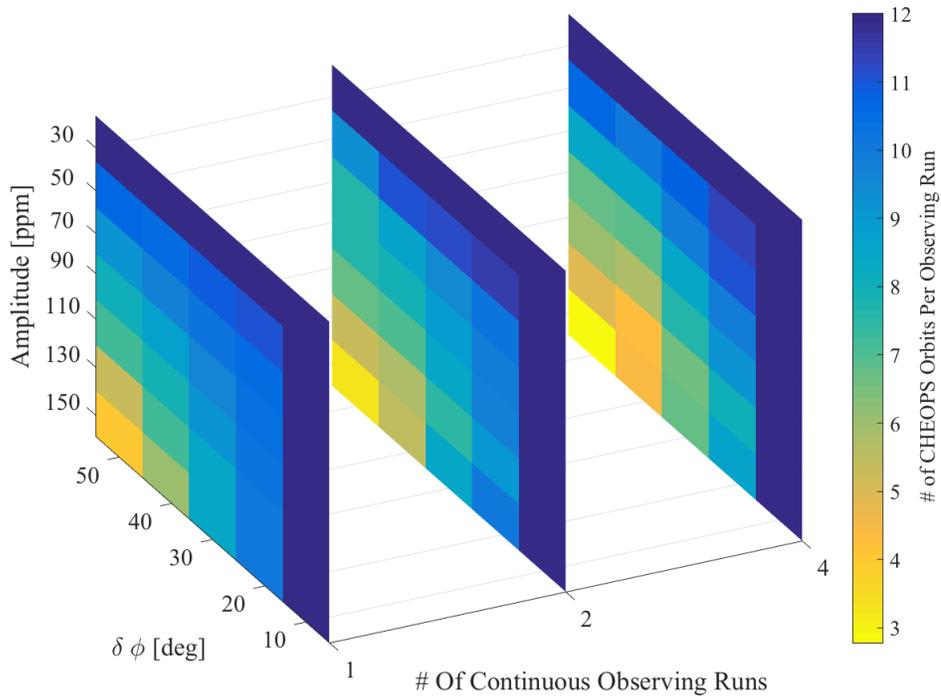

Fig. 11.— Bayes' factor maps resulting from simulated CHEOPS observations of an ultra-short period planet. Colors represent the number of CHEOPS orbits per observing run necessary for model $M_{\text{offset}}$ to be strongly favored over the model $M_{\text{no offset}}$. Note that the assumed orbital period is 0.8d, which is fully covered with 12 CHEOPS orbits. Therefore phase curves with amplitudes less than 10ppm or offsets less than $30^o$ will not be detectable with anything less than the full phase curve.